\documentclass[twocolumn]{article}

\usepackage{graphicx} % Required for including images
\usepackage[utf8]{inputenc} % For special characters
\usepackage[T1]{fontenc} % Encoding
\usepackage{times} % Times New Roman font
\usepackage[letterpaper, margin=1in]{geometry} % Margins and paper size US Letter
\usepackage{setspace} % For line spacing
\usepackage{cite} % For citation
\usepackage{caption} % For caption formatting
\usepackage{subcaption} % For sub-figures
\usepackage{amsmath,amsfonts,amssymb} % For math formatting and symbols
\usepackage{listings}
\usepackage[font=it]{caption}
\usepackage{hyperref}
\usepackage{url}

\title{Emergence of autopoietic vesicles able to grow, repair and reproduce in a minimalist particle system}
\author{Thomas Cabaret\\
	\small Nice, France\\
	\small Email: thomas.cabaret@gmail.com}
\date{\today}

\begin{document}
	
	\maketitle
	
	\begin{abstract}
		This paper describes a 2D particle system in which autopoietic vesicles that are able to grow, repair, and reproduce can emerge from an initial random distribution of individual components. The reproduction process also exhibits inheritance of the vesicle content that could in principle change from one generation to the next and influence fitness, making those autopoietic vesicles eligible to be acted upon by natural selection.
		Video abstract: \href{https://www.youtube.com/watch?v=JmCN_4jTwf8}{link}.
	\end{abstract}
	
	\section*{Keywords}
	Origin of life, Simulation, Particle system, Autopoiesis, Autocatalytic set, Abiogenesis, Artificial chemistry, A-Life.
	
	\section{Introduction}
	The primary motivation of this research is to simulate an origin-of-life type event that exhibits properties we consider essential. These properties initially included:
	\begin{itemize}
		\item Conservation of matter \textit{(conservation of the number of particles)}
		\item Aspects of non-Darwinian evolution of structures akin to chemical evolution \textit{(progression towards the structures of interest prior to their actual formation)}
		\item Emergent structures with a flexible dynamic that, in principle, could be acted upon by natural selection
	\end{itemize}
	The simulation is not intended to replicate the exact physical or chemical conditions of the real world but seeks to be sufficiently analogous to reproduce key concepts or phenomena associated with an origin-of-life process. This objective has guided our choice to utilize the continuous physics of a particle system.

	Using artificial environments to witness some form of origin-of-life related events has been the subject or prior works in the cellular automata world\cite{ChouRef}\cite{Evoloop}, or with particle systems \cite{PPSRef}\cite{Squirm3}\cite{DigiHive}. \cite{ChouRef} is a variation of the Langton's loops showing spontaneous emergence of replicators from an initial random distribution, and a limited form of evolution. It comes with highly handcrafted rules and bypasses important real world constraints like conservation of matter. Evoloop\cite{Evoloop} is another variation of Langton's loop that stresses a richer evolvability, at the expense of losing spontaneous formation capability. On the particle system side, PPS\cite{PPSRef} highlights the relationship between rulesets and emergence of replicators. While it aligns more with real-world constraints, the spontaneously formed structures currently lack a clear mechanism for evolution and extending the system towards this goal would be challenging. Conversely, Squirm3\cite{Squirm3} or DigiHive\cite{DigiHive}, other particle systems in the realm of artificial chemistry\cite{ArtificialChemistry}, exhibit highly ordered structures able to reproduce with information inheritance from one generation to the next. Unfortunately, these structures cannot form spontaneously, which limits their application in the specific context of our research.
	
	It should also be noted that physical systems based on fatty acid vesicles have been successfully constructed to exhibit either some form of spontaneous formation or a dynamic of growth and division\cite{PrimitiveCellularCompartments}\cite{Protocell}.
	
	Our 2D particle system aims to integrate spontaneous formation with reproductive dynamics. The governing rules ensure matter conservation and lead to the emergence of vesicular structures capable of growth, self-repair, and reproduction. Additionally, our system tries to find the right balance between fine-tuning and emergent behavior, concentrating on the stability and higher-order properties of these emergent replicators. These include their ability to transmit some form of information from one generation to the next, thereby making them potentially subject to evolutionary processes.

	\section{Definitions}
    Prior to describing our particle system and corresponding findings, it is essential to introduce two fundamental concepts that underpin its construction, autocatalytic sets and autopoiesis, which play important roles in various other origin of life research.
	
%	\subsection{Autocatalytic sets}
	We will refer to any set of chemical reactants as an \textbf{autocatalytic set} if, for any reactants \textit{(catalysts included)} not provided by the environment, there is a reaction within the set that produces those reactants and ensures their population increases over time.
	\begin{figure}[h]
	\centering
	\includegraphics[width=1.0\linewidth]{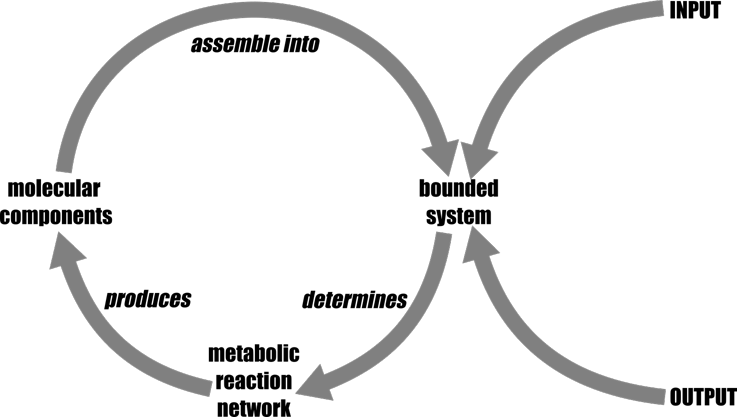}
	\caption{Living system as an autopoietic structure, adapted from \cite{AutopoRef}}
	\label{fig:Autopoietic}
    \end{figure}

%	\subsection{Autopoiesis}
	A structure will be said \textbf{autopoietic} when it is able to produce its components and repair itself from within.
	See \textbf{figure \ref{fig:Autopoietic}} from \textit{(Luisi 2003)}\cite{AutopoRef} where a living system can be understood as the alliance of structure and inner workings, such that the structure enables the inner workings, and the inner workings, in turn, sustain the structure. In addition of using autocatalysis, autopoietic structures are also able to maintain their organization and integrity. This concept was initially introduced by Varela and Maturana \cite{AutopoiesisCognition} in the early 1980s and has been regularly refined since. We consider it pivotal in the emergence of life, as it may serve as a precursor to reproduction. Simple autopoietic structures often exhibit the ability to grow through the same mechanism they use for self-maintenance. Large ones may divide when subjected to physical constraints that cause them to break apart. The inherent autopoietic capabilities then facilitate the repair of these fragments, effectively completing a full reproduction cycle.
	
%   Should be here to be on page 3
	\begin{figure*}[p]
	\centering
	\includegraphics[width=\textwidth]{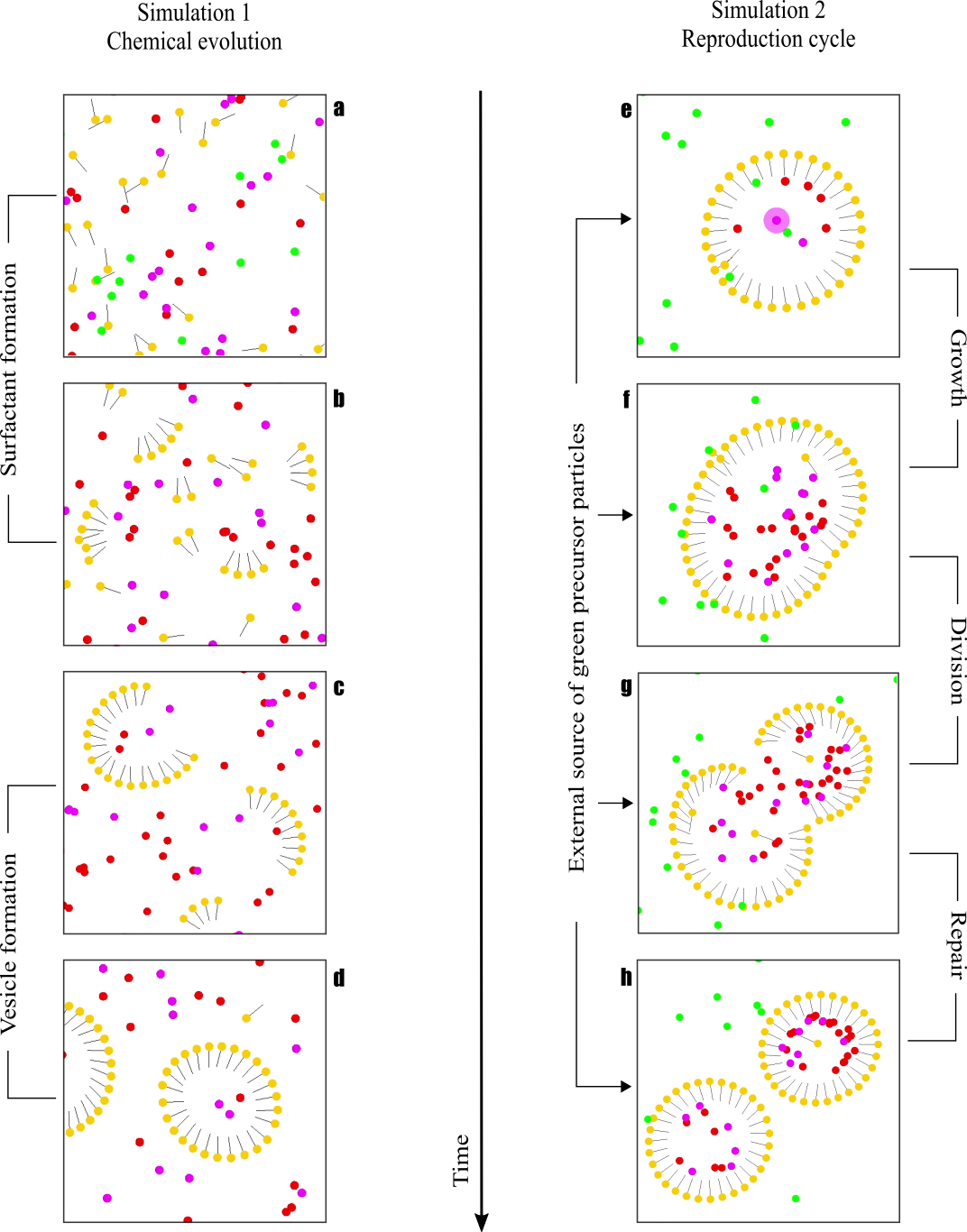}
	\caption{Simulation 1, chemical evolution: starting from a random distribution of individual components and leading to non empty vesicle formation. Simulation 2: starting from a fully formed autopoietic vesicle exposed to a source of precursor particles and leading to the vesicle reproduction. \href{https://www.youtube.com/watch?v=JmCN_4jTwf8}{video link}}
	\label{fig:simuStages}
	\end{figure*}
	
	\section{Simulation and findings}
    In order to gain insight into the particle system we have developed and the significant dynamics exhibited in various simulations, we suggest to start by examining one such simulation and provide a cursory explanation of the different rules governing our particle system as we progress \textit{(This simulation is available as a YouTube video see annex 2)}. A more exhaustive and technical presentation of these rules will be reserved for the next chapter.
  
    We will start a simulation from a random distribution of particles and describe the various stages of chemical evolution, in addition to highlighting the characteristics of autonomously formed structures able to grow, repair, and reproduce.
	\begin{table*}[h]
	\centering
	\hspace*{-0.7cm}
	\begin{tabular}{|c|c|c|c|c|c|}
		\hline
		& \textbf{Precursor} & \textbf{Catalyst} & \textbf{Intermediary} & \textbf{Surfactant} \\
		\hline
		\textbf{Precursor} & none & only chemical & only chemical & none \\
		\hline
		\textbf{Catalyst} & only chemical & none & none & close range repulsion \\
		\hline
		\textbf{Intermediary} & only chemical & none & none & close range repulsion \\
		\hline
		\textbf{Surfactant} & none & close range repulsion & close range repulsion & complex force profile, see text for details \\
		\hline
	\end{tabular}
	\caption{Summary of interactions between particle types.}
	\label{tab:forcesTable}
    \end{table*}
	In \textbf{figure \ref{fig:simuStages}a} we see such an initial random distribution of the 4 types of particles used in our model: green particles will be referred to as \textit{Precursors}, red particles will be referred to as \textit{Catalysts}, purple particles as \textit{Intermediary}, and yellow particles as \textit{Surfactant} or polar particles. After a short while, like illustrated by \textbf{figure \ref{fig:simuStages}bc}, we see small aggregates of yellow particles that exhibit an intrinsic curvature. On \textbf{figure \ref{fig:simuStages}d} we see that when enough surfactant particles aggregate full vesicles are eventually formed, some of which can entrap either at least a catalyst or an intermediary particle. A second simulation, starting from a fully formed vesicle, shows that those vesicles containing catalysts are indeed autopoietic structures able to grow, repair, and reproduce. When such a vesicle is placed close to a source of precursor particles, those can, by design, cross the surfactant membrane, and a proto-metabolic activity can start inside, allowed by a set of chemical rules changing types of particles when specific ones come close to each other. The net effect of the inner chemical reactions is to consume precursors and increase the inner population of catalyst, intermediary and surfactants particles that eventually aggregate to the membrane from within, making the whole structure grow. See \textbf{figure \ref{fig:simuStages}ef}. Large vesicles are unstable, start to squirm and eventually break apart into 2 halves that initially stay close to each other, as illustrated in \textbf{figure \ref{fig:simuStages}g}. This proximity usually prevents the leaking of a significant part of the content. An autopoietic activity persists in both halves, producing surfactants that leads quickly to both vesicle repair, achieving a full reproduction cycle. See \textbf{figure \ref{fig:simuStages}h}.
	
	\section{Model and genesis}
\begin{figure}[ht]
	\centering
	\includegraphics[width=0.7\linewidth]{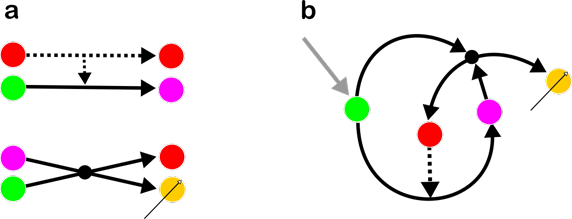}
	\caption{a) The two chemical rules of our particle system. b) The minimalist autocatalytic set they define. Bold arrows for chemical transformations, dashed arrows for catalytic activity and gray arrows for external intake. One color dot by particle type: precursor(green), catalyst(red), intermediary(purple) and surfactant(yellow).}
	\label{fig:simuChemicalRules}
    \end{figure}

	Our particle system involves two chemical rules, see \textbf{figure \ref{fig:simuChemicalRules}a}: A Precursor can be transmuted into an Intermediary particle in presence of a Catalyst, and if a Precursor and an Intermediary particle interacts they are both transmuted into Catalyst and Surfactant. This set of chemical rules allows the existence of a minimalist autocatalytic set, see \textbf{figure \ref{fig:simuChemicalRules}b}. In the simulation both of those chemical rules are triggered if eligible particles are within an interaction radius, small enough to prevent most of the chemical interactions between particles located on opposite sides of a surfactant membrane.
    
\begin{figure}[ht]
		\centering
		\includegraphics[width=0.7\linewidth]{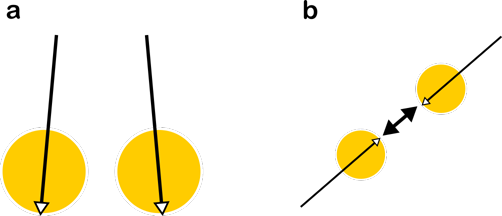}
		\caption{Surfactant molecules force profile details a) equilibrium state, b) repulsive domain.}
		\label{fig:forceS}
	\end{figure}

	Our particle system also involves forces between particles that can modify their trajectories. The force profile between particles is summarized in table \ref{tab:forcesTable}.
	The only non trivial force profile concerns surfactant molecules, it’s carefully handcrafted to lead to a satisfactory vesicular behavior. The key idea to understand this force profile is to think of it as derived from a potential with minimum being reached when surfactant molecules are close to each other, side by side, tail on the same side, almost parallel to each other with a slight angle. See \textbf{figure \ref{fig:forceS}a}.
	
	If a couple of interacting yellow particles are far from this equilibrium state, each will experience a force towards it. The slight deviation from perfect parallelism in this stable equilibrium state \textit{(figure \ref{fig:forceS}a)} is a way to impose intrinsic local curvature to surfactant molecule aggregates. This deviation being constant, too large vesicles are by nature unstable as all of their surfactant molecules cannot be at equilibrium with their neighbors simultaneously. It was designed that way on purpose with the hope it would lead to some kind of division and the first behavior observed was satisfactory, as seen previously, large vesicles break apart and then repair. The force profile of surfactant molecules also contain a repulsive domain when they are close, head to head, and anti-parallel, see \textbf{figure \ref{fig:forceS}b}. The key idea of this specificity is to avoid fusion of vesicles freshly formed as at the interaction location between two vesicles, surfactant molecules will be in this repulsive domain, making fully formed vesicles repulsive to one another which strongly favors the behavior towards division rather than fusion.
	\begin{figure}[ht]
		\centering
		\includegraphics[width=1.0\linewidth]{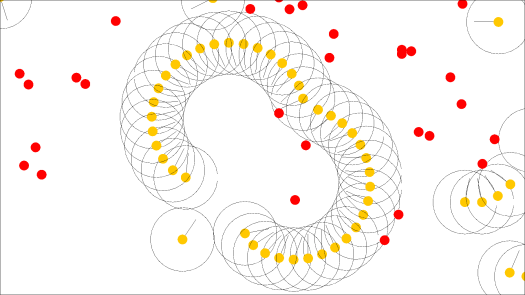}
		\caption{Visualization of force interaction radius illustrated by the black circles around surfactant particles.}
		\label{fig:interactionRadius}
	\end{figure}
	All the force interaction radius are small, as illustrated by the circles in \textbf{figure \ref{fig:interactionRadius}} showing interaction radius of surfactant particles which barely allow each of them to see the neighbor of their neighbor within a vesicle. It is particularly noteworthy that the interaction range is substantially smaller than the size of the structures of interest, namely the vesicles. This significant difference in scale between local interactions and the overall structure justifies to talk about emergence when referring to the vesicles and their dynamic.

\begin{figure}[h]
		\centering
		\includegraphics[width=1.0\linewidth]{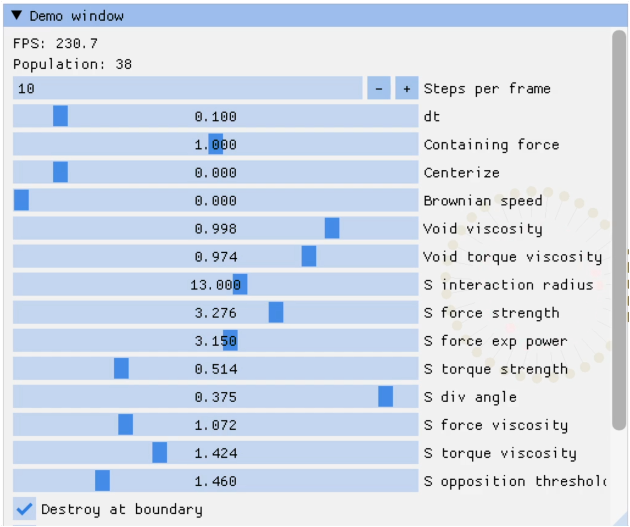}
		\caption{User Interface illustrating various controls on the simulation parameters.}
		\label{fig:UI}
	\end{figure}

	Experimentation with the simulation parameters and fine-tuning were necessary. To facilitate this process, a minimalist user interface was designed, see \textbf{figure \ref{fig:UI}}, featuring slider controls that allow for the adjustment of various parameters and observation of their effects on the behaviors of vesicles.
	
	We can then conclude the description of our particle system with additional remarks
	\begin{itemize}
		\item Energy is not conserved: the model contains dissipative terms that are unavoidable to observe structure formation without simulating solvent particles. The model also includes the exact opposite, Brownian random noise to favor macroscopic motion of particles and thus particle encounter.
		\item Momentum is conserved (up to numerical approximations).
		\item Angular momentum not conserved: the force profile of surfactant molecules can in some cases lead them to spiral one toward the other to end up side by side.
	\end{itemize}
	For more details about the model, please refer to the c++ code in annex.
	
	\section{Discussion}
	Despite not having gathered specific metric on the following points, various manually started simulations show a high stability of vesicle formation from an initial random distribution: with an initial state involving several hundreds of particles it is common to see half a dozen fully formed and non empty vesicles thus making this period of chemical evolution highly reproducible.
	
	The reproduction seems also quite resilient, although it can sometimes fail, by example through leakage of the whole content, success seems more frequent than failure, and most importantly it almost never happens that a reproduction failure leads to the loss of both daughter vesicles, most of the times at least one remains from which a reproduction can be retried. In average the reproduction success rate seems compatible with an exponential growth of the vesicle population, it is quite common to witness several successful divisions happening sequentially.
	
	A limitation of our particle system is the need to split the simulation into two distinct ones: the first for vesicles formation, the second for vesicles reproduction. In the current state of the model, if a source of precursor particles would be added into the initial random starting distribution, those precursors would mostly react outside emergent vesicles before reaching the interior of one of them and allowing a more interesting activity. This limitation could be addressed by allowing larger simulations that incorporate local macroscopic variations in the environment, akin to real-world scenarios such as distinct conditions in a pool at one location versus an ocean elsewhere. Introducing time-varying conditions to simulate real-life cycles could achieve the same goal. Yet another solution would be to adjust the chemical rules to ensure a minimal concentration of precursor particles outside vesicles. 
	
	During various experiments it appeared quite clear that the key for a lot of relevant behaviors is the existence of the membrane itself. Having particles able to aggregate into vesicles with a flexible enough dynamic is essential to define macroscopic systems, to unite unrelated entities into a single one. Although membrane formation can emerge from numerous sets of rules, stepping among one of them by pure luck and still have a computationally efficient model seems unlikely to us. We suggest to study this kind of particle system, in which vesicular behavior is manually crafted to study higher order properties, as the dedicated class of \textit{vesicular particle systems}, which seems to be justified by the unique macroscopic dynamic they allow. We also wish to highlight that much of the current model's value lies in the natural appearance and behavior of the simulation. The rules and mechanisms driving the emergent processes are minimalist, yet they enable smooth chemical evolution and reproduction. This is a primary objective of such research; assessing the simplicity of crafting the ruleset and the authenticity of the simulation provides valuable insights into plausible timelines for the origin-of-life event in the real world.
	
	Finally our simulation could be chemically improved by introducing other types of particles allowing in principle the evolution of our vesicles as it is discussed in the next chapter.

	\section{Evolvability of Autopoietic Vesicles}
    A pivotal question in the study of structures pertains to their alignment with Darwinian principles: Are such structures susceptible to natural selection? This essentially involves demonstrating that:
	\begin{itemize}
		\item Some information is inherited.
		\item This information can undergo variations from one generation to another.
		\item Such variations can influence fitness.
	\end{itemize}
	The inheritance mechanism of our vesicles is quite straightforward: their content is conveyed from one generation to the next: the set of types of particles is conserved with a high probability, and to a lesser extent the relative concentration between types of particles. If persistent variations in vesicle content occur from one generation to the next, this could lead to changes in the inner chemical activity, thereby impacting fitness through numerous mechanisms, as observed in real-world living cells. Currently, this is not feasible with our simplified model version. However, by just enriching the chemistry rules of the simulation, it might be possible to enable at least one of the following two mechanisms to facilitate some form of evolution.
	
	\subsection{Autocatalytic extension}
	The most natural extension of our current model involves incorporating additional chemical rules beyond the existing four particle types and two reactions. By doing so, we aim to simulate a dynamic resembling natural selection in which the autocatalytic network itself would act as the repository of information passed from one generation to the next as it is understood in the context of RAF theory\cite{RAFTheory}. Upon vesicle division, the chemical content is distributed between the two daughter cells. Depending on the chemical species in focus, one of two scenarios may unfold:
	\begin{itemize}
	\item If a chemical species is not autocatalytically produced by the network, or if its production rate is insufficient relative to the division rate, a progressive dilution and eventual disappearance of this species will occur across several reproductive cycles.
	\item Conversely, if a chemical species is autocatalytically produced at an adequate rate, it will persist within the network, continuing to participate in internal reactions and be part of the inherited information from one generation to the next.
    \end{itemize}
	We can highlight that the former dilution dynamic is a type non-Darwinian evolution, where initially cluttered vesicles self-purify, retaining only those chemicals that participate in the autocatalytic network.
	
\begin{figure}[ht]
	\centering
	\includegraphics[width=1.0\linewidth]{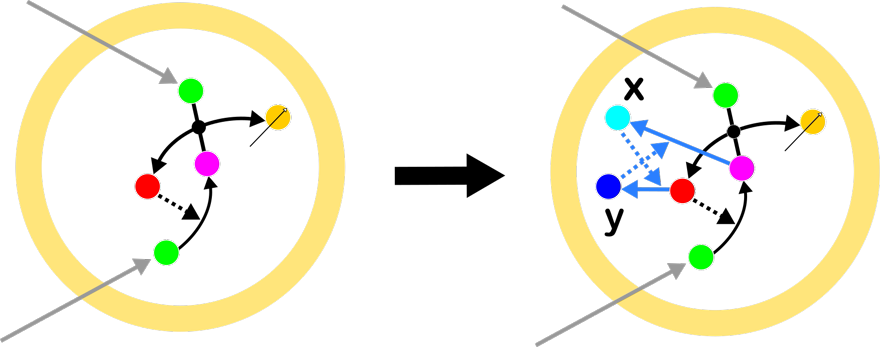}
	\caption{Example of change by autocatalytic extension: addition of new x, y types of particles within a vesicle can lead to a new autocatalytic set able to sustain itself from one generation to the next. Bold arrows for chemical transformations, dashed arrows for catalytic activity and gray arrows for external intake. One color dot by particle type: precursor(green), catalyst(red), intermediary(purple) and surfactant(yellow).}
	\label{fig:autocatalyticExtention}
	\end{figure}

\begin{figure}[h]
	\centering
	\includegraphics[width=0.7\linewidth]{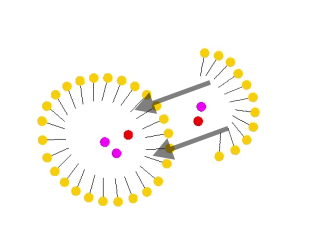}
	\caption{Mechanism for vesicle entry discovered during simulation (illustration): outer particles can end up within vesicle by being trapped between membrane chunks during their fusion with vesicles.}
	\label{fig:smartLeakage}
    \end{figure}

	A possibility for the content to change across generations would simply be serendipitous leakage within a vesicle. For instance, during certain simulations, we observed a novel mechanism whereby new particles enter a vesicle, becoming trapped between the vesicle and an aggregating small membrane fragment, as depicted in \textbf{figure \ref{fig:smartLeakage}}. For a variation to be sustained within the lineage, and so be considered as a true mutation in the context of Darwinian evolution, it should consist of a set of particles that we categorize as "autocatalytic extensions," an example is given in \textbf{figure \ref{fig:autocatalyticExtention}}, where both types of added particles are generated autocatalytically from chemical species present in the network of the previous generation. Related concepts have been discussed and further theorized by Vasas, V., Fernando, C., Santos, M., et al. in \cite{BeforeGenes}.

	Regarding the link between the internal chemical network and fitness, one can rely on the canonical relationship where variations in the internal networks influence fitness through the rate of surfactant particle synthesis, thereby affecting the division rate.
	
	The primary disadvantage of this suggested model improvement is its reliance on rare events to consistently alter the chemical network and the limited open-endedness of the evolution it allows. However, its strength relies on the ease of implementation as an enhancement to our current model and the fact it is supported by prior analysis \cite{BeforeGenes}.


	\subsection{Programmable reactants}
	A subsequent improvement to foster an evolution dynamic of interest would be to introduce RNA-like behavior. This final chapter is more speculative but outlines how we could proceed with subsequent models.
	
\begin{figure}[h!]
	\centering
	\includegraphics[width=1.0\linewidth]{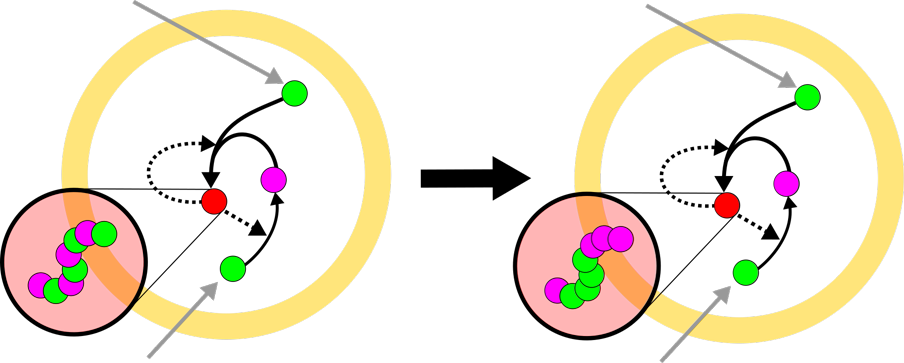}
	\caption{Example of change by error in template copy. In this figure, the catalyst in red is a macroscopic abstraction of a polymer made of green and purple particles. It catalyzes its own formation by canonical template copy. Bold arrows for chemical transformations, dashed arrows for catalytic activity and gray arrows for external intake. One color dot by particle type: precursor(green), catalyst(red), intermediary(purple) and surfactant(yellow).}
	\label{fig:programmableEvolution}
	\end{figure}

	The particles in our simulation are currently more analogous to real-world molecules than mere atoms. Consequently, a feasible enhancement would be to refine the granularity of the current model to accommodate polymers composed of various smaller particles. The rule set would be adapted to allow some of these polymers to catalyze their own formation through a standard template-copy mechanism. Template-copyable polymers within the metabolic network, as shown in \textbf{figure \ref{fig:programmableEvolution}}, would act as their own catalysts and could be considered programmable, since they retain their autocatalytic properties despite variations in their sequence, while these alterations simultaneously influence their functional role in the chemical network. Developing the ruleset that governs template-copy mechanisms and correlates sequence with chemical function is undoubtedly a challenging endeavor. While technically demanding, this task leans more towards the practical rather than the conceptual side. Notably, the successful implementation of template-copy simulations has already been achieved \cite{TemplateCopySimulation}, thus reducing the challenge to a matter of integrating these mechanisms into a cohesive model.

	With the introduction of such a new ruleset and, over an evolutionary timeline, the encapsulation of such template-copyable catalysts could be regarded as an evolution of evolvability. Before this encapsulation, the only mutation pathway available is through the catalytic extension mechanism described in the previous chapter, which depends on rare environmental events and offers limited open-endedness. With a progressive inclusion of programmable polymers into proto-metabolism, variations through template copying errors would become possible, thus enhancing the potential for variation independently of external factors.

	In summary, this aligns with a 'metabolism-first' approach, and we consider it theoretically feasible to extend the model to witness an evolution timeline as follows:
	\begin{enumerate}
	\item Formation of autopoietic vesicles,
	\item Purification of autocatalytic content,
	\item Selection of an optimal catalytic network (not open-ended)
	\item Progressive replacement of some chemical species by increasingly ideal template copyable catalysts
	\item Selection based on the chemical efficacy of the best polymers. (increase in open-endedness)
	\end{enumerate}

	To realize this significant advancement over the current model, we recommend utilizing a more mature framework of artificial chemistry, such as Christian Heinemann’s Alien-project\cite{AlienProject}, an optimized and versatile simulation environment for such implementations.

	\section{Conclusion}
	In this paper, we have presented a 2D particle system that exhibits emergent formation of autopoietic vesicles capable of growth, repair, and reproduction. We have shown that it is possible for structures with these properties to emerge from a simple set of rules governing particle interactions. We highlighted that these structures exhibit inheritance of their content and discussed how the system could be extended to allow relevant changes from one generation to the next, thus allowing those structure to be acted upon by natural selection.
    While our simulation is highly simplified compared to real-world chemical systems, it serves as a proof-of-concept that such behaviors are possible by giving an new A-life example of a minimalist origin of life event. Our model provides a foundation for further exploration of these concepts and could be extended and refined to better mimic complex chemistry and to explore other potential mechanisms for the emergence and evolution of autopoietic structures. We believe that meticulously crafting rules to achieve a pre-visualized timeline of evolution provides profound insights into what is both natural and plausible as a conceptual framework for the real-world origin of life.
	
	\section*{Acknowledgments}
	Thank to Thomas Schmickl who suggested the idea to keep track of this personal work on arXiv.
	
	\appendix
	\section{Appendix}
	\subsection{Program Code}
	Please refer to github repository:
	\url{https://github.com/ThomasCabaret/SimulationSkeleton/tree/arxiv}
 	\textit{(use the arxiv branch)}, the readme contains the pieces of information to reproduce both experiments.
	\subsection{Demo}
	A condensed trailer version of our particle system is available on YouTube:
	\url{https://www.youtube.com/watch?v=JmCN_4jTwf8}


\begin{thebibliography}{99}
		
	\bibitem{ChouRef}
	Chou H.H., Reggia J.A. \textit{Emergence of self-replicating structures in a cellular automata space.}
	Physica D: Nonlinear Phenomena, 110, 252-276 (1997).
	\url{https://doi.org/10.1016/S0167-2789(97)00132-2}
	
	\bibitem{Evoloop}
	Sayama H. \textit{A new structurally dissolvable self-reproducing loop evolving in a simple cellular automata space.} Artificial Life, 5(4), 343-65 (1999). 
	\url{https://doi.org/10.1162/106454699568818}
	
	\bibitem{PPSRef}
	Schmickl T., Stefanec M., Crailsheim K. \textit{How a life-like system emerges from a simplistic particle motion law.} Scientific Reports 6, 37969 (2016). 
	\url{https://doi.org/10.1038/srep37969}
	
	\bibitem{Squirm3}
	Hutton T.J. \textit{Evolvable self-reproducing cells in a two-dimensional artificial chemistry.} Artificial Life, 13(1), 11-30 (2007). 
	\url{https://doi.org/10.1162/artl.2007.13.1.11}
	
	\bibitem{AutopoRef}
	Luisi P.L. \textit{Autopoiesis: a review and a reappraisal.} Naturwissenschaften 90, 49–59 (2003). 
	\url{https://doi.org/10.1007/s00114-002-0389-9}
	
	\bibitem{AutopoiesisCognition}
	Maturana H. R., Varela F. J. \textit{Autopoiesis and Cognition: The Realization of the Living.} Dordrecht, Netherlands: D. Reidel Publishing Co., 1980.
	\url{https://link.springer.com/book/10.1007/978-94-009-8947-4}
	
	\bibitem{DigiHive}
	Sienkiewicz R., Jędruch W. \textit{DigiHive: Artificial Chemistry Environment for Modeling of Self-Organization Phenomena} Artificial Life (2023) 29 (2): 235–260. (2023).
	\url{https://doi.org/10.1162/artl_a_00398}
	
	\bibitem{PrimitiveCellularCompartments}
	Hanczyc M. M., Fujikawa S. M., Szostak J. W. \textit{Experimental Models of Primitive Cellular Compartments: Encapsulation, Growth, and Division.} Science (2003) 302 (5645): 618–622.
	\url{https://doi.org/10.1126/science.1089904}
	
	\bibitem{Protocell}
	Toparlak Ö. D., Sebastianelli L., Ortuno V. E., Karki M., Xing Y., Szostak J. W., Krishnamurthy R., Mansy S. S. \textit{Cyclophospholipids Enable a Protocellular Life Cycle.} ACS Nano 2023 17 (23), 23772-23783
	\url{https://pubs.acs.org/doi/10.1021/acsnano.3c07706}
	
	\bibitem{ArtificialChemistry}
	Dittrich P., Ziegler J., Banzhaf W. \textit{Artificial Chemistries A Review.} Artificial Life (2001) 7 (3): 225–275. 
	\url{https://doi.org/10.1162/106454601753238636}

	\bibitem{BeforeGenes}
	Vasas, V., Fernando, C., Santos, M. et al. \textit{Evolution before genes.} Biol Direct (2012) 7: 1. \url{https://doi.org/10.1186/1745-6150-7-1}

	\bibitem{TemplateCopySimulation}
	Mori, T., Kawamata, I., Murata, S. \textit{Self-replication and Mutation of Polymeric Molecules Simulated by Simplified Chemistry}, 2022 Tenth International Symposium on Computing and Networking Workshops (CANDARW), Himeji, Japan, 2022, pp. 192-198. \url{https://doi.org/10.1109/CANDARW57323.2022.00083}

	\bibitem{RAFTheory}
	Hordijk, W. \textit{Evolution of Autocatalytic Sets in Computational Models of Chemical Reaction Networks.} Orig Life Evol Biosph (2016) 46: 233–245. \url{https://doi.org/10.1007/s11084-015-9471-0}
	
	\bibitem{AlienProject}
	Heinemann C. \textit{Alien Project.}
	\url{https://alien-project.org/}
	% Add more references as needed
	
	\end{thebibliography}
\end{document}